\documentclass{article}
\pdfoutput=1
\usepackage{jheppub}

\newcommand\fft[2]{{\frac{#1}{#2}}}
\newcommand\ft[2]{{\textstyle\frac{#1}{#2}}}
\newcommand\nn{\nonumber}

\renewcommand{\Re}{\operatorname{Re}}
\DeclareMathOperator{\Tr}{Tr}
\DeclareMathOperator{\csch}{csch}

\begin{document}

\preprint{LCTP-19-33}

\title{Subleading corrections to the free energy in a theory with $N^{5/3}$ scaling}

\author{James T. Liu}
\author{and Yifan Lu}
\affiliation{Leinweber Center for Theoretical Physics, Randall Laboratory of Physics,\\
The University of Michigan, Ann Arbor, MI 48109-1040, USA}

\emailAdd{jimliu@umich.edu}
\emailAdd{yifanlu@umich.edu}

\abstract{We numerically investigate the sphere partition function of a  Chern-Simons-matter theory with $SU(N)$ gauge group at level $k$ coupled to three adjoint chiral multiplets that is dual to massive IIA theory.  Beyond the leading order $N^{5/3}$ behavior of the free energy, we find numerical evidence for a term of the form $(2/9)\log N-(1/18)\log k$.  We conjecture that the $(2/9)\log N$ term may be universal in theories with $N^{5/3}$ scaling in the large-$N$ limit with the Chern-Simons level $k$ held fixed.}

\maketitle \flushbottom

\section{Introduction}

Over the last few years, remarkable progress has been made in our understanding of supersymmetric partition functions and precision tests of AdS/CFT beyond leading order in the large-$N$ expansion.  In addition to super-Yang-Mills theories in four dimensions, there is much interest in three dimensional Chern-Simons-matter theories generalizing ABJM theory \cite{Aharony:2008ug}.  Such theories generally fall into two classes, the first being ABJM-like and with $N^{3/2}$ scaling, where the sum of Chern-Simons levels $k_a$ of the gauge groups vanishes, and the second, with $N^{5/3}$ scaling when the sum does not.  While ABJM-like theories have been extensively studied, less is known about those with $N^{5/3}$ scaling.  The aim of the present paper is to explore the structure of subleading corrections in such theories through a numerical evaluation of the sphere partition function in a particularly simple model.

From a holographic point of view, the sum of Chern-Simons levels is related to the Romans mass, $2\pi l_sF_0=\sum_ak_a$ \cite{Gaiotto:2009mv,Gaiotto:2009yz}, and hence ABJM-like theories with $F_0=0$ can often be associated with M-theory duals.  Remarkably, the sphere partition function for such theories takes the form of an Airy function \cite{Marino:2009jd,Drukker:2010nc,Herzog:2010hf,Drukker:2011zy,Fuji:2011km,Marino:2011eh}.  Expansion in the M-theory limit then immediately gives the structure of free energy beyond the leading order
\begin{equation}
    F(N,k_a)=f_0(k_a)N^{3/2}+f_1(k_a)N^{1/2}+\fft14\log N+f_2(k_a)+\mathcal O(N^{-1/2}),
\end{equation}
where we take $F=-\log Z$.  In general, the coefficients are model dependent.  However, the coefficient of $\log N$ is universal, and can be reproduced exactly by a one-loop computation in the dual supergravity on AdS$_4\times X_7$ \cite{Bhattacharyya:2012ye}.  Similarly, the topologically twisted index for ABJM theory \cite{Benini:2015noa,Benini:2015eyy}, which has been used to count the microstates of BPS black holes in AdS$_4$, has a universal (ie independent of chemical potentials) subleading $\log N$ contribution \cite{Liu:2017vll,PandoZayas:2019hdb} that can be reproduced from a one-loop computation in eleven-dimensional supergravity \cite{Jeon:2017aif,Liu:2017vbl}.

Here we extend some of the numerical investigations into the case of theories with $N^{5/3}$ scaling.  In particular, we consider $\mathcal N=2$ Chern-Simons gauge theory with gauge group $SU(N)$ at level $k$ coupled to three adjoint chiral multiplets (denoted $X$, $Y$ and $Z$) and with superpotential $W=\Tr X[Y,Z]$.  This theory was first described in \cite{Guarino:2015jca} as the dual to a particular compactification of massive IIA theory on AdS$_4\times S^6$.  The topologically twisted index for this theory was used for black hole microstate counting in \cite{Hosseini:2017fjo,Benini:2017oxt,Azzurli:2017kxo} and studied numerically in \cite{Liu:2018bac}.  In this case, the subleading structure has the form
\begin{equation}
    \Re\log Z(N,k=1)=f_0N^{5/3}+f_1N^{2/3}+f_2N^{1/3}-\fft7{18}\log N+f_3+\mathcal O(N^{-1/3}),
\end{equation}
where again the $\log N$ term was numerically observed to be universal.

In this paper, we focus on the sphere partition function of the same model and obtain numerical evidence for an expansion of the form
\begin{align}
    \Re F(N,k)&=f_0N^{5/3}k^{1/3}+\left(\fft12\log2\pi-1\right)N+f_1N^{2/3}k^{4/3}+f_2N^{1/3}k^{-1/3}\nn\\
    &\quad+\fft29\log N-\fft1{18}\log k+f_3(k)+\mathcal O(N^{-1/3}).
    \label{eq:spf}
\end{align}
The structure of this subleading expansion is similar to that of the topologically twisted index, although numerically we find an additional term of $\mathcal O(N)$ which we argue is an artifact of the saddle point approximation that we employ.  We have also explored the free energy in the 't~Hooft limit and found
\begin{align}
    \Re F(N,\lambda)&=N^2(f_0\lambda^{-1/3}+f_1\lambda^{-4/3})+\left(\fft12\log2\pi-1\right)N+\fft16\log N\nn\\
    &\quad+\left(f_2\lambda^{1/3}+\fft1{18}\log\lambda+\tilde f_3(\lambda)\right)+\mathcal O(N^{-1}),
\end{align}
which is compatible with the expansion in the M-theory limit.  Note that the $(1/6)\log N$ term is a contribution to the exact partition function that is not visible in the genus expansion.

Although our main results are obtained numerically, we provide partial analytic support for the structure of the sub-leading terms in the free energy.  In particular, within the framework of the saddle point expansion, we demonstrate that the leading term includes a $-(2/3)N\log N$ contribution which is, however, canceled by an equal but opposite contribution from the one-loop determinant.  This term arises from the log divergent short distance behavior as adjacent eigenvalues approach each other, and is also present in the individual Bethe potential and Jacobian determinant components of the corresponding topologically twisted index \cite{Liu:2018bac}.

In the next section, we briefly review the sphere partition function for the model we are considering and summarize its leading order behavior.  We then highlight the results of the numerical investigation, including the determination of the $(2/9)\log N$ term, in section~\ref{sec:numerical}.  In section~\ref{sec:structure}, we provide a partial justification of the form of the expansion, (\ref{eq:spf}).  Finally, we conclude in section~\ref{sec:discussion} with a conjecture on the universality of the log corrections to the sphere partition function.

\section{Leading order free energy in the dual of massive IIA string theory}

We are interested in the sphere partition function for the $\mathcal N=2$ Chern-Simons-matter theory presented in \cite{Guarino:2015jca}.  This theory has gauge group $SU(N)_k$ and three adjoint chiral multiplets, and its partition function can be obtained via localization \cite{Kapustin:2009kz,Jafferis:2010un,Hama:2010av,Guarino:2015jca}, with the result
\begin{equation}
Z=\int \prod_{i=1}^{N} \frac{d \lambda_i}{2 \pi} \prod_{i<j}^{N} \left(4 \sinh^2\left(\frac{\lambda_i-\lambda_j}{2}\right)\right) \exp\left(3\sum_{i,j}^N\ell\left(\frac{1}{3}+\frac{i}{2 \pi}(\lambda_i-\lambda_j)\right)+\frac{ik}{4\pi}\sum_i^N \lambda_i^2\right),
\label{eq:Zexact}
\end{equation}
where the function $\ell(z)$ arises from the one-loop matter determinant and satisfies $\partial_z\ell(z)=-\pi z\cot(\pi z)$ and can be integrated to give \cite{Jafferis:2010un,Hama:2010av}
\begin{equation}
\ell(z)=-z \log(1-e^{2\pi i z})+\fft{i}2\left(\pi z^2+\fft1\pi\mathrm{Li}_2(e^{2\pi i z})\right)-\fft{i\pi}{12}.
\end{equation}

Before examining the higher-order corrections to the sphere free energy, we first review the leading order result \cite{Guarino:2015jca}.  In the large-$N$ limit, it is natural to make a saddle point approximation.  The solution to the saddle point equations is generally complex, so we take $\lambda=N^{\alpha}(x+i y(x))$ where $x$ is real and $y(x)$ is a real function.  Here we have assumed that the eigenvalues scale with $N$ with exponent $\alpha$.  In the large-$N$ limit, the distribution of $x$ and $y$ becomes dense and we use $\rho(x)$ to describe the density of the real part of the eigenvalues. Then, at leading order, the saddle-point approximation to (\ref{eq:Zexact}) gives $Z=e^{-S}$ where the effective action takes the form \cite{Herzog:2010hf,Jafferis:2011zi,Guarino:2015jca},
\begin{equation}
S=\fft{N^{1+2 \alpha}}{4\pi} k \int dx \rho(x) (2xy(x)-i(x^2-y^2))+\fft{16}{27} \pi^2 N^{2-\alpha}\int dx \frac{\rho^2(x)}{1+i y^\prime (x)}.
\label{eq:S0}
\end{equation}
In order to obtain a non-trivial solution for the saddle point, we require that both terms scale similarly in $N$, and this determines $\alpha=1/3$.

The leading order free energy can be obtained by extremization of the effective action subject to the normalization constraint $\int dx \rho(x)=1$.  This can be performed by introducing a Lagrange multiplier $\mu$ and adding a contribution
\begin{equation}
    \delta S=-\mu\left(\int dx \rho(x)-1\right),
\end{equation}
to (\ref{eq:S0}).  Varying with respect to $y(x)$ and $\rho(x)$ and normalizing the eigenvalue density then gives the leading-order result
\begin{equation}
y_0(x)=\fft1{\sqrt{3}}x,\quad \rho_0(x)=\fft3{4x_*}\left(1-\left(x/x_*\right)^2\right),\quad\mbox{where}\quad x_*=\fft{2^{2/3}\pi}{3^{1/6}k^{1/3}}.
\label{eq:leading}
\end{equation}
Finally, inserting this solution into the effective action, and using the convention $F=-\log Z$, gives the leading order behavior of the $S^3$ free energy \cite{Guarino:2015jca}
\begin{equation}
F=e^{-i\pi/6}\fft{2^{4/3}\pi}{5\cdot3^{1/3}}k^{1/3}N^{5/3}\qquad\Rightarrow\qquad
\Re F=\fft{2^{1/3}\cdot3^{1/6}\pi}5 k^{1/3} N^{5/3},
\label{eq:N53}
\end{equation}
which has the expected $N^{5/3}$ scaling.

\section{Numerical investigation of the free energy}
\label{sec:numerical}

While the large-$N$ results are straightforward to obtain, the higher order contributions have proven to be a challenge to obtain analytically.  Thus, to provide guidance on the structure of the higher order terms, we turn to a numerical investigation.  Note that, unlike the cases where there is a Bethe ansatz like approach, such as the topologically twisted index on $\Sigma_g\times T^n$ \cite{Benini:2015noa,Benini:2016hjo} or the rewriting of the $S^3\times S^1$ partition function in a Bethe ansatz form \cite{Benini:2018mlo}, here the exact partition function (\ref{eq:Zexact}) involves integrals over the matrix eigenvalues $\lambda_i$.

Instead of performing these integrals numerically, we limit our investigation to the large-$N$ limit and the saddle-point expansion.    Note, however, that the Chern-Simons-matter theory is governed by two parameters, $N$ and $k$, and there are complementary ways of taking the large-$N$ limit.  The natural IIA expansion of (\ref{eq:Zexact}) corresponds to the genus expansion
\begin{equation}
    F(g_s,t)=\sum_{g=0}^\infty g_s^{2g-2}F_g(t)=\fft1{g_s^2}F_0(t)+F_1(t)+\cdots,
    \label{eq:genus}
\end{equation}
were $g_s=2\pi i/k$ and the 't~Hooft coupling $t=g_sN=2\pi iN/k$ is held fixed.  Here, $F_0(t)$ is the leading-order saddle point term, and $F_1(t)$ is evaluated by the one-loop determinant.  On the other hand, one can also consider the M-theory expansion where the Chern-Simons level $k$ is held fixed.  Here, the numerical expansion takes the form
\begin{equation}
    F(N,k)=\mathcal F_0(N,k)+\mathcal F_1(N,k)+\cdots,
\end{equation}
where $F_0(N,k)$ is the saddle point contribution at fixed $k$ and $F_1(N,k)$ arises from the Gaussian determinant around the saddle point.

In principle, both expansions ought to be equivalent.  However it is well known that there are non-perturbative effects (such as worldsheet and membrane instantons) that may not be visible in one or the other expansion \cite{Ooguri:2002gx,Marino:2011eh,Hanada:2012si,Hatsuda:2012dt,Calvo:2012du,Hatsuda:2013oxa}.  Of course, numerically, we only evaluate the partition function for finite $N$ and $k$ (and only up to the Gaussian determinant).  Nevertheless, we can probe either the `t~Hooft limit or the M-theory limit by holding either $N/k$ fixed or $k$ fixed when extrapolating to large~$N$.  We mainly focus on the M-theory limit, although we have also compared our numerical results with those obtained by holding $t$ fixed.

\begin{figure}[t]
\centering
\includegraphics[width=0.46\textwidth]{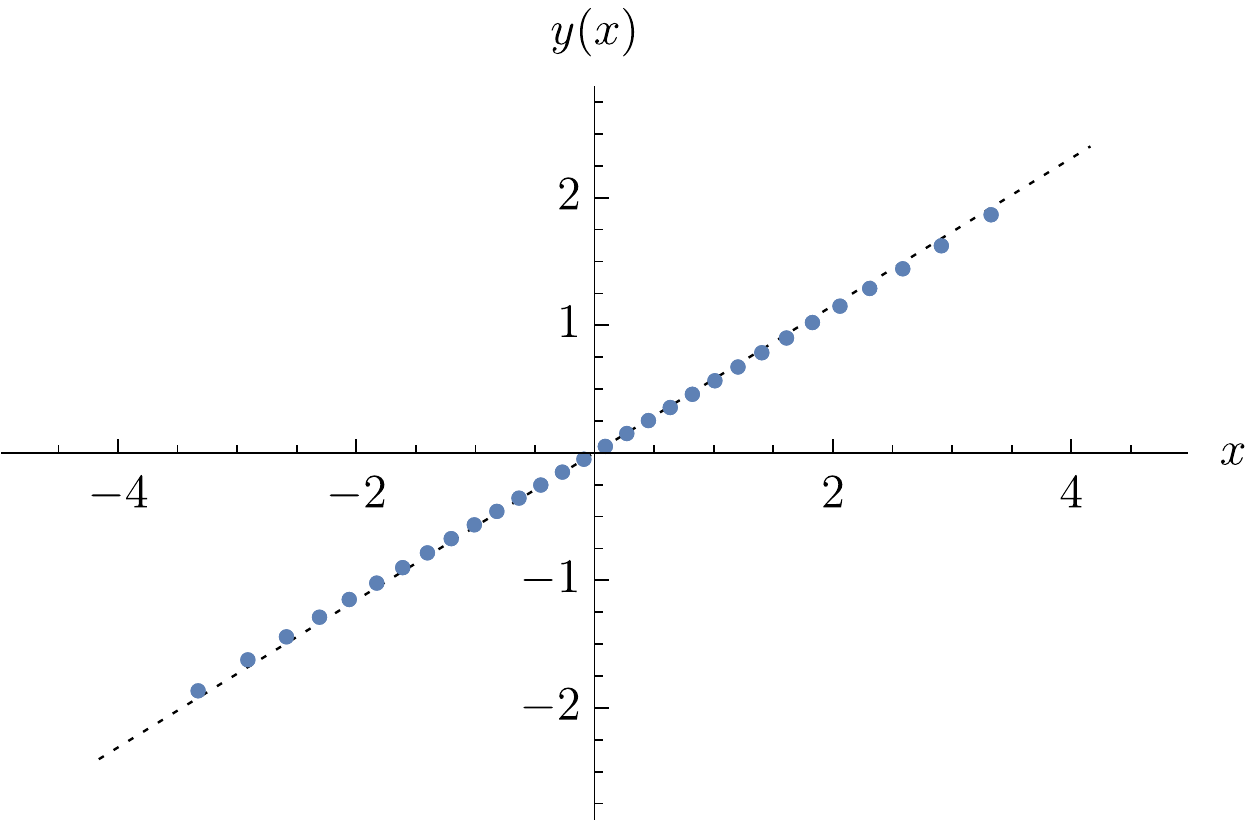}
\includegraphics[width=0.46\textwidth]{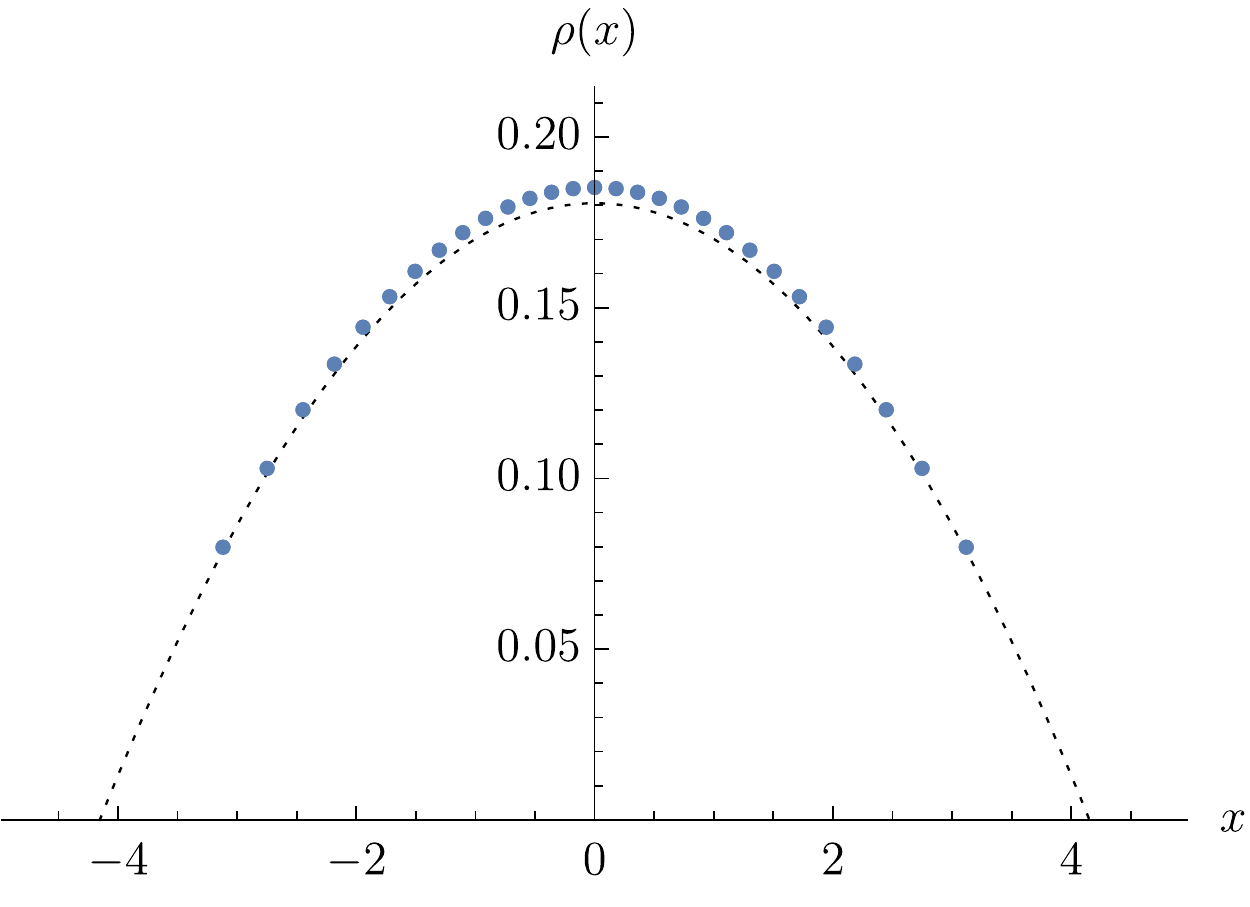}
\caption{Numerical solution for the eigenvalues $\lambda_i=N^{1/3}(x_i+iy_i)$ (left) and eigenvalue density $\rho(x)$ (right) for $N=30$ and $k=1$.  The leading order solution, (\ref{eq:leading}), is shown by the dotted line.}
\label{fig:N=30}
\end{figure}

The first term in the expansion of the free energy comes directly from the partition function (\ref{eq:Zexact})
\begin{equation}
\mathcal F_0(N,k;\lambda_i)=-\fft{ik}{4\pi}\sum_{i} \lambda_i^2-\sum_{i<j}\log\left(4 \sinh^2\left(\fft{\lambda_i-\lambda_j}{2}\right)\right)-\sum_{i,j} 3\ell\left(\fft13+\fft{i}{2 \pi}(\lambda_i-\lambda_j)\right).
\label{eq:calF0}
\end{equation}
The eigenvalues $\lambda_i$ are determined by solving the saddle point equations
\begin{equation}
\fft{\partial\mathcal F_0}{\partial \lambda_i}=-\fft{ik}{2\pi}\lambda_i-\sum_{j\ne i} \coth\left(\fft{\lambda_i-\lambda_j}{2}\right)+\sum_{j} \frac{2 \sinh(\lambda_i-\lambda_j)-\fft{3 \sqrt 3}{2\pi}(\lambda_i-\lambda_j)}{1+2 \cosh(\lambda_i-\lambda_j)} = 0.
\label{eq:saddle}
\end{equation}
We use Mathematica, and in particular the built-in FindRoot function, to solve these equations numerically.  For a given value of $N$ and $k$, FindRoot is first called with WorkingPrecision set to MachinePrecision, and with an initial set of eigenvalues determined by the large-$N$ distribution, (\ref{eq:leading}).  The solution is then refined with a second call to FindRoot with WorkingPrecision set to 100.  All solutions are checked for convergence before evaluation of the free energy.  An example of a generated eigenvalue distribution is shown in Figure~\ref{fig:N=30}.  Although we work with a range of $N$ from 100 to 600 in steps of 20, the figure is presented with $N=30$ and $k=1$ to highlight the discrete nature of the eigenvalues and its deviation from the leading-order large-$N$ solution.  (The eigenvalue density $\rho(x)$ is obtained by taking finite differences.)

\begin{figure}[t]
\centering
\includegraphics[width=0.46\textwidth]{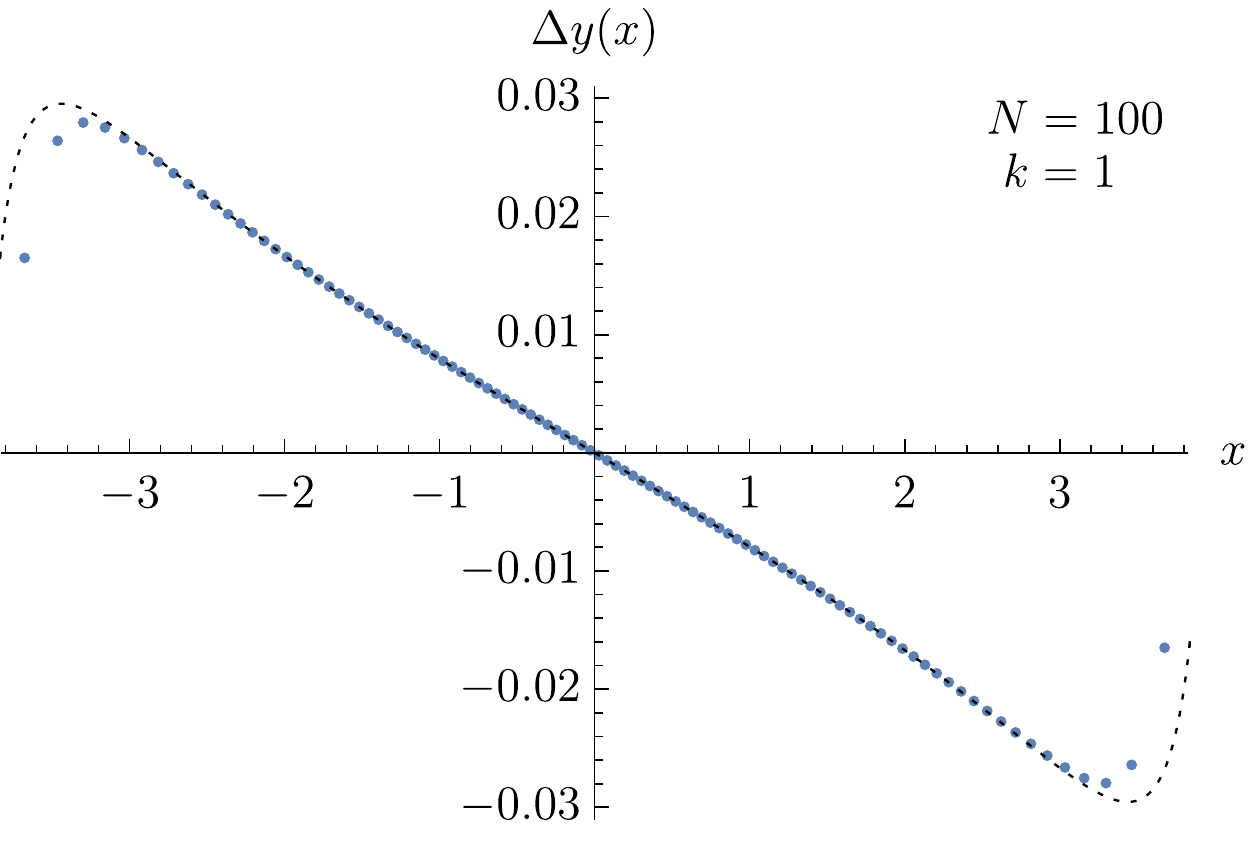}
\includegraphics[width=0.46\textwidth]{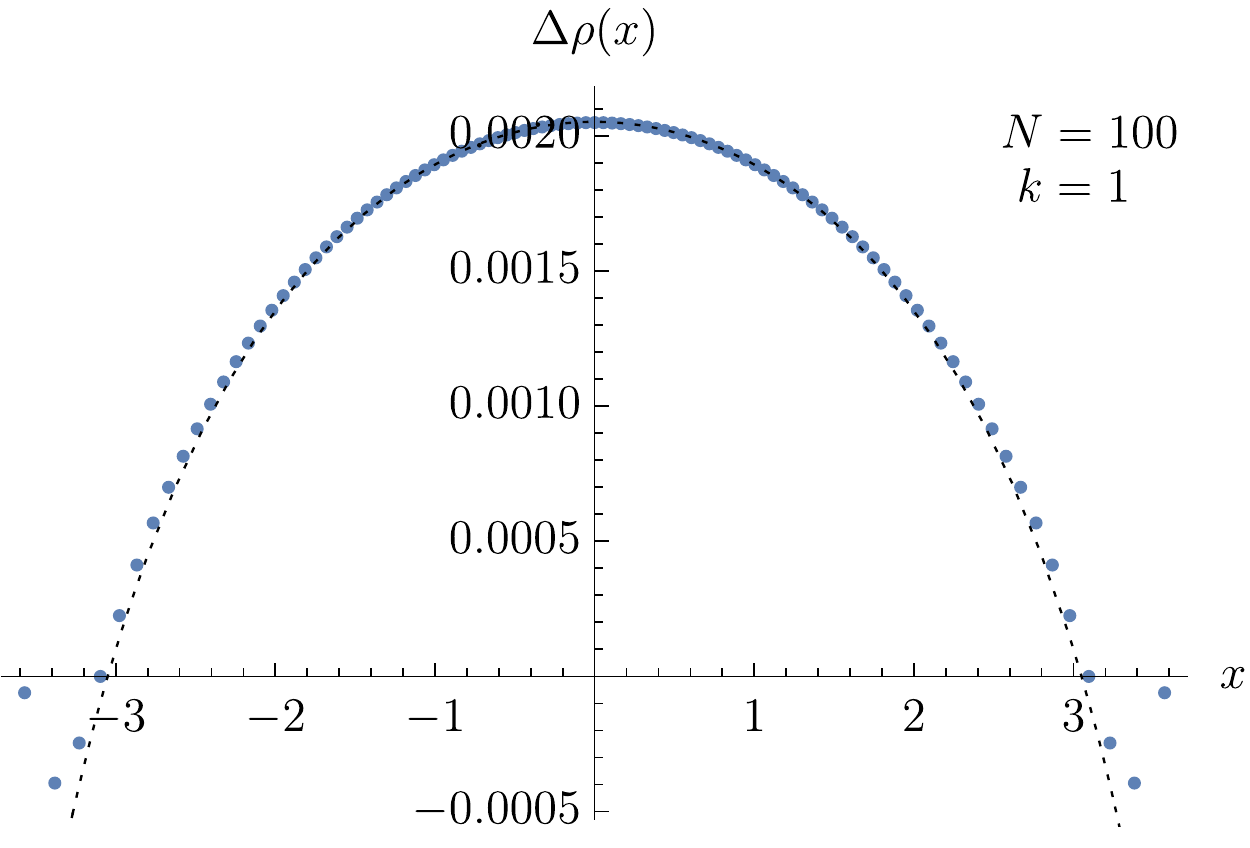}
\caption{The numerical solution for $y(x)$ and $\rho(x)$ with the leading order behavior subtracted out.  The dotted lines correspond to the subleading solution, (\ref{eq:rho1}), with the unknown constants determined numerically. Here we have taken $N=100$ and $k=1$.}
\label{fig:rhosub}
\end{figure}

\begin{figure}[t]
\centering
\includegraphics[width=0.7\textwidth]{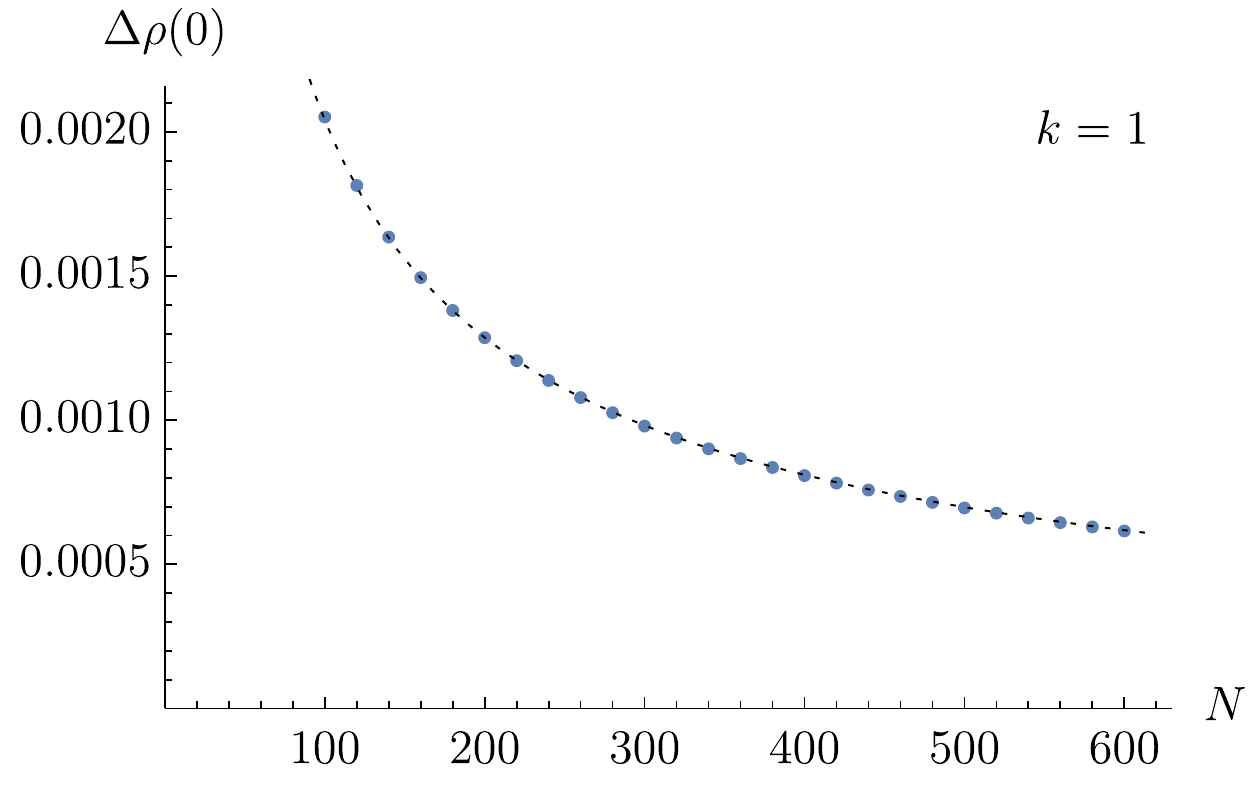}
\caption{The difference $\Delta\rho(0)$ as a function of $N$ for $k=1$.  The dotted line is the numerical fit $\Delta\rho(0)=0.0440/N^{2/3}$.}
\label{fig:rho0}
\end{figure}

As can be seen from the numerical $N=30$ solution, the eigenvalues deviate somewhat from the leading order solution.  To get a sense of the higher order corrections, we can examine the differences $\Delta y(x)=y(x)-y_0(x)$ and $\Delta\rho(x)=\rho(x)-\rho_0(x)$ where the leading order functions $y_0(x)$ and $\rho_0(x)$ are given in (\ref{eq:leading}). An example of the subleading behavior is given in Figure~\ref{fig:rhosub} for $N=100$ and $k=1$.  To be somewhat more quantitative, we plot the difference $\Delta\rho(0)$ at the midpoint of the distribution as a function of $N$ in Figure~\ref{fig:rho0}.  A fit to the numerical data demonstrates that the first subleading correction scales as $\mathcal O(N^{-2/3})$. This result will be useful in guiding our analytic approximations below.

Once the eigenvalues are determined numerically from the saddle point equation, (\ref{eq:saddle}), they may be directly inserted into the expression (\ref{eq:calF0}) for $\mathcal F_0$.  We also compute the one-loop determinant contribution
\begin{equation}
\mathcal F_1(N,k;\lambda_i)=\frac{1}{2}\log \det\left(\fft{\partial^2\mathcal F_0}{\partial\lambda_i\partial\lambda_j}\right)+\fft{N}2\log2\pi,
\label{eq:F1det}
\end{equation}
and evaluate the free energy at the level of $\mathcal F_0+\mathcal F_1$. Note that the factor $(N/2)\log2\pi$ arises for each eigenvalue from a combination of $\sqrt{2\pi}$ from the Gaussian integral and $1/2\pi$ from the normalization of the integration region in (\ref{eq:Zexact}).  In addition, we only consider the real part of the free energy, as there are potential branch issues leading to $2\pi i$ ambiguities in the numerical evaluation of $F=-\log Z$.

At leading order, the numerical data reproduces the $\mathcal O(N^{5/3})$ behavior, (\ref{eq:N53}), very well, so we naturally subtract it out to highlight the subleading corrections.  The next term we find 
is linear in $N$, and has a coefficient that is numerically very close to $(1/2)\log2\pi-1$.  This leads us to conjecture that it is in fact a precise match, and we remove this term as well before fitting for the remaining subleading corrections.  This is justified \textit{a posteriori} by the quality of the fits that we obtain without use of a linear $N$ term.

Numerically, we take integer values of $k$ from 1 to 7.  For a fixed $k$, we then generate data for $N=100$ to $600$ in steps of $20$ and perform a linear least squares fit to the expansion
\begin{equation}
F(N,k)=f_0(k)N^{5/3}+\left(\fft12\log2\pi-1\right)N+c_1(k) N^{2/3}+c_2(k) N^{1/3}+c_3(k)\log N+c_4(k)+\sum_{i=1}^5\frac{d_i(k)}{N^{i/3}}.
\label{eq:F(N,k)}
\end{equation}
Our main interest is in the coefficients $c_i(k)$ of terms that do not vanish in the $N\to\infty$ limit.  However, for fitting purposes, we include a set of terms that scale as $1/N$ to some power in order to account for higher order terms in the expansion of the free energy.  We do not expect the $d_i(k)$ coefficients to be numerically reliable, although their magnitudes tend to be of order unity so they are under reasonable control.  The fit coefficients are displayed in Table~\ref{tbl:fixedk}.

\begin{table}[t]
\centering
\begin{tabular}{l|rrrr}
$k$&$c_1$&$c_2$&$c_3$&$c_4$\\
\hline
1&$-0.01108$&$-0.61301$&$0.22247$&$0.69630$\\
2&$-0.02793$&$-0.48662$&$0.22342$&$0.64921$\\
3&$-0.04795$&$-0.42514$&$0.22390$&$0.62474$\\
4&$-0.07037$&$-0.38630$&$0.22423$&$0.61010$\\
5&$-0.09476$&$-0.35863$&$0.22446$&$0.60140$\\
6&$-0.12083$&$-0.33751$&$0.22465$&$0.59699$\\
7&$-0.14841$&$-0.32062$&$0.22482$&$0.59587$
\end{tabular}
\caption{The numerical fit for the coefficients of $\{N^{2/3},N^{1/3},\log N,1\}$ in the saddle point evaluation of the free energy.  The fit is performed independently for each fixed value of $k$ with $N$ from 100 to 600.}
\label{tbl:fixedk}
\end{table}

A quick glance at Table~\ref{tbl:fixedk} suggests that the coefficient $c_3$ of the $\log N$ term is nearly constant, although it increases slightly with $k$.  Assuming this is a numerical artifact, we are led to conjecture that $c_3=2/9$ exactly.  This is in line with other examples where the coefficient of the $\log N$ term is known either exactly or numerically to be a simple rational number.  The other coefficients in Table~\ref{tbl:fixedk} are more obviously $k$-dependent.  However, a numerical fit suggests that the coefficient $c_1$ of the $N^{2/3}$ term scales exactly as $k^{4/3}$ and likewise that the coefficient $c_2$ of the $N^{1/3}$ term scales as $k^{-1/3}$, both with small residuals.  This leads us to conjecture the large-$N$ but fixed $k$ expression for the free energy
\begin{equation}
    \Re F(N,k)=f_0(k)N^{5/3}+\left(\fft12\log2\pi-1\right)N-0.01108N^{2/3}k^{4/3}-0.61315N^{1/3}k^{-1/3}+\fft29\log N+\mathcal O(1),
    \label{eq:freek}
\end{equation}
where the numerical coefficients are obtained by a least squares fit to $c_1(k)=\bar c_1k^{4/3}$ and $c_2(k)=\bar c_2k^{-1/3}$, respectively.

So far, we have examined the free energy in the large-$N$ limit while holding $k$ fixed.  In contrast, the `t~Hooft limit is taken by holding the `t~Hooft coupling $t=2\pi iN/k$ fixed.  Note that, although the Chern-Simons level $k$ is integer quantized, the numerical solution to the saddle point equations, (\ref{eq:saddle}), and hence the numerical free energy can be obtained for arbitrary real values of $k$.  This allows us to more directly examine the genus expansion of the free energy.  For convenience, we remove the factor of $2\pi i$ from the `t~Hooft coupling, and define $\lambda=N/k$.  We can then compute the free energy numerically following the procedure outlined above with $N=100$ to $600$ in steps of $20$, but this time holding $\lambda$ fixed from $50$ to $300$ in steps of $50$.  A least squares fit for the free energy then gives
\begin{align}
    \Re F(N,\lambda)&=N^2\left(\fft{2^{1/3}\cdot 3^{1/6}\pi}5\lambda^{-1/3}-0.01104\lambda^{-4/3}\right)+\left(\fft12\log2\pi-1\right)N\nn\\
    &\quad+\fft16\log N+\fft1{18}\log\lambda-0.612\lambda^{1/3}+\cdots,
    \label{eq:freelam}
\end{align}
where there is numerical uncertainty in the last digit of the final term. Note that the numerical coefficients match those in the fixed $k$ expansion, (\ref{eq:freek}), provided we take $k=N/\lambda$.

At this point, several comments are in order.  Firstly, we always subtract the known leading order behavior, which in this case corresponds to the $N^2\lambda^{-1/3}$ term.  Secondly, for each fixed value of $\lambda$, a numerical fit is performed to a function composed of integer powers of $N$ from $N^2$ down to $N^{-3}$.  After this, the coefficients of each monomial are fitted as a function of $\lambda$.  Finally, we have written down analytic coefficients for the $\log N$ and $\log\lambda$ terms.  The coefficient of $\log N$ was initially obtained numerically by including such a term in the linear least squares fit.  Since the resulting fit was numerically close to $1/6$ we conjectured that it is precisely this value. Making this compatible with the $2/9$ factor in the fixed $k$ expansion, (\ref{eq:freek}) then demands the addition of the $(1/18)\log\lambda$ term.  With this conjecture, the analytic terms are in fact subtracted out before fitting for the numerical coefficients in (\ref{eq:freelam}).

The expression (\ref{eq:freelam}) for the free energy is naturally organized according to the genus expansion, (\ref{eq:genus}), which can be rewritten as
\begin{equation}
    F(N,\lambda)=N^2F_0(\lambda)+F_1(\lambda)+\cdots.
\end{equation}
In particular, we find
\begin{equation}
    \Re F_0(\lambda)=\fft{2^{1/3}\cdot 3^{1/6}\pi}5\lambda^{-1/3}-0.01104\lambda^{-4/3},\qquad\Re F_1(\lambda)=\fft1{18}\log\lambda-0.612\lambda^{1/3}+\cdots,
\end{equation}
where numerically we find no additional terms in $F_0(\lambda)$ but are less certain about $F_1(\lambda)$.  Note, however, that we find an additional contribution
\begin{equation}
    \Re\tilde F=\left(\fft12\log2\pi-1\right)N+\fft16\log N,
\end{equation}
which is not captured by the genus expansion.

The term linear in $N$ is partially analytical, with the $\log2\pi$ factor arising directly from the various $2\pi$ factors in the partition function and normalization of the Gaussian measure.  Curiously, however, the $-1$ factor is only obtained numerically, and arises from a combination of the leading-order $\mathcal F_0$ and one-loop determinant $\mathcal F_1$.  Support for this sort of combination at the linear-$N$ level will be seen below when we turn to an analytic investigation of the terms.  Nevertheless, we expect that the overall linear-$N$ term is most likely an artifact of the saddle point expansion, as it, for example, is not present in the topologically twisted index, which can be evaluated exactly (up to numerical precision) as a sum over Bethe roots \cite{Liu:2018bac}.

\section{The structure of the \texorpdfstring{large-$N$}{large-N} expansion}
\label{sec:structure}

As we have seen numerically, with $k$ held fixed the large-$N$ free energy receives subleading corrections with various powers of $N$. We now take a closer look at the structure of the large-$N$ expansion and provide support for the numerical fitting function that was used in (\ref{eq:F(N,k)}).  The starting point is of course the matrix partition function (\ref{eq:Zexact}), which we write as
\begin{equation}
    Z(N,k)=\int\prod_{i=1}^N\fft{d\lambda_i}{2\pi}e^{-\mathcal F_0(N,k;\lambda_i)},
\end{equation}
where
\begin{equation}
    \mathcal F_0(N,k;\lambda_i)=-\fft{ik}{4\pi}\sum_i^N\lambda_i^2-\sum_{i,j}^Nf(\lambda_i-\lambda_j),
    \label{eq:Sact}
\end{equation}
with
\begin{equation}
    f(z)=\fft12\log(4\sinh^2(z/2))+3\ell(\ft13+\ft{i}{2\pi}z).
    \label{eq:f(z)}
\end{equation}
Note that the log term is divergent for $z=0$ and should not be included in the sum when $i=j$.

We take the large-$N$ limit by assuming the eigenvalues condense on a single cut and then converting the sums into integrals using the Euler-Maclaurin formula
\begin{align}
    \sum_{i=1}^Nf_i&=\int_1^Ndi\,f(i)+\fft12\bigl(f(N)+f(1)\bigr)+\fft1{12}\bigl(f'(N)-f'(1)\bigr)+\cdots,\nn\\
    &=(N-1)\int_{x_1}^{x_2}\rho(x)dx\,f(x)+\fft12\bigl(f(x_2)+f(x_1)\bigr)+\fft1{12(N-1)}\left(\fft{f'(x_2)}{\rho(x_2)}-\fft{f'(x_1)}{\rho(x_1)}\right)+\cdots,
\end{align}
where we have introduced the eigenvalue density $di=(N-1)\rho(x)dx$.  Note that this provides a formal $1/N$ expansion of the action $S(N,k)$, even though its saddle point value is only associated with genus zero in the `t~Hooft expansion.

The first term in the action, (\ref{eq:Sact}), is easily dealt with, and we find
\begin{equation}
    S_1=-\fft{ik}{4\pi}N^{1+2\alpha}\left[\left(1-\fft1N\right)\int_{-x_*}^{x_*} dx\,\rho(x)(x+iy(x))^2+\fft1N(x_*+iy(x_*))^2+\mathcal O(1/N^2)\right],
    \label{eq:S1}
\end{equation}
where we have made the substitution
\begin{equation}
    \lambda_i\to\lambda(x)=N^\alpha(x+iy(x)).
    \label{eq:lambdai}
\end{equation}
Although we always take $\alpha=1/3$, we prefer to keep it in these expressions to highlight the nature of the expansion both in powers of $1/N$ from the genus expansion and Euler-Maclaurin terms and in powers of $1/N^\alpha$ from the large `t~Hooft parameter $\lambda=N/k$ limit.  Note that we assume the eigenvalues are symmetrically distributed in the interval $x\in[-x_*,x_*]$ with $y(x)$ an odd function of $x$.

The second term in (\ref{eq:Sact}) is a bit more delicate as we must handle the log divergence of the function $f(z)$.  Although this is excluded from the discrete sum, in the large-$N$ limit the eigenvalues become dense and hence $\lambda_i-\lambda_j$ becomes vanishingly small for $i$ close to $j$.  One way to handle this is to introduce a regulated $f(z)$ function
\begin{equation}
    \tilde f(i,j)=\begin{cases}f(\lambda_i-\lambda_j)-\log(|i-j|\alpha(j)),&i\ne j;\\3\ell(\fft13+\fft{i}{2\pi}z),&i=j,\end{cases}
\end{equation}
where
\begin{equation}
    \alpha(j)=\fft{j-1}{N-1}(\lambda_j-\lambda_{j-1})+\fft{N-j}{N-1}(\lambda_{j+1}-\lambda_j)
\end{equation}
is an interpolated difference of adjacent eigenvalues that remains valid at the endpoints.  We now have
\begin{equation}
    S_2=-\sum_{i,j}^N\tilde f(i,j)-\sum_{i\ne j}^N\log(|i-j|\alpha(j)).
    \label{eq:S2reg}
\end{equation}
The first sum is taken over $i$ and $j$ without restriction as the regulated $\tilde f(i,j)$ is well behaved even when $i$ approaches $j$.  Note that the regulator $\log(|i-j|\alpha_j)$ grows logarithmically for $i$ well separated from $j$, so it cannot be ignored.  However, the sum over $\log|i-j|$ can be performed to yield
\begin{equation}
    S_2=-\sum_{i,j}^N\tilde f(i,j)-(N-1)\sum_j^N\log\alpha(j)-2\log G(N+1),
    \label{eq:S2sum}
\end{equation}
where $G(N+1)$ is the Barnes $G$ function.

At this stage, the two sums in (\ref{eq:S2sum}) can be converted to integrals through Euler-Maclaurin summation.  Working only to the first non-trivial order, we obtain
\begin{align}
    S_2&=-\int_1^Ndi\int_1^Ndj\tilde f(i,j)-\fft12\int_1^Ndi\bigl(\tilde f(i,1)+\tilde f(1,i)+\tilde f(i,N)+\tilde f(N,i)\bigr)\nn\\
    &\quad-(N-1)\left(\int_1^Ndi\log\alpha(i)+\fft12\bigl(\log\alpha(1)+\log\alpha(N)\bigr)\right)-2\log G(N+1).
    \label{eq:S2int}
\end{align}
Although it was important to work with the regulated function $\tilde f(i,j)$ when converting the first sum into an integral, now that the expression is written as an integral, we can split $\tilde f(i,j)$ back into its original and regulator components since log divergences can be integrated.  Integrating the regulator then gives a result which nearly cancels the second line of (\ref{eq:S2int}).  However, the cancellation is not perfect, and we are left with
\begin{align}
     S_2&=-\int_1^Ndi\int_1^Ndj f(i,j)-\fft12\int_1^Ndi\bigl(f(i,1)+ f(1,i)+ f(i,N)+ f(N,i)\bigr)+\int_1^Ndi\log\left(\fft{\alpha(i)}{2\pi}\right),
\end{align}
up to terms of $\mathcal O(1)$.  The log term that shows up here is essentially a result of transforming the sum of a log divergent expression into an integral.

We now convert the integrals over $i$ and $j$ into integrals along the cut where the eigenvalues condense.  Along with the replacement $di=(N-1)\rho(x)dx$, we also need an expression for $\alpha(i)$, which can be obtained in the continuum limit as
\begin{equation}
    \alpha(i)=\fft{d\lambda}{di}=\fft{d\lambda/dx}{(N-1)\rho(x)}=\fft{N^\alpha(1+iy'(x))}{(N-1)\rho(x)},
    \label{eq:alphai}
\end{equation}
where we made use of (\ref{eq:lambdai}).  As a result, we find
\begin{align}
    S_2&=-N^2\biggl[\left(1-\fft2N\right)\int_{-x_*}^{x_*}dx\int_{-x_*}^{x_*}d\tilde x\,\rho(x)\rho(\tilde x)f(x,\tilde x)+\fft1N\int_{-x_*}^{x_*}dx\,\rho(x)\bigl(f(x,x_*)+f(x_*,x)\bigr)\nn\\
    &\kern4em+\fft1N\int_{-x_*}^{x_*}dx\,\rho(x)\log\left(\fft{2\pi\rho(x)}{1+iy'(x)}\right)\biggr]-(1-\alpha)N\log N+\mathcal O(1),
    \label{eq:S2N}
\end{align}
where
\begin{equation}
    f(x,\tilde x)=f(N^{\alpha}((x-\tilde x)+i(y(x)-y(\tilde x)))),
    \label{eq:fxxt}
\end{equation}
and $f(z)$ was defined in (\ref{eq:f(z)}).

So far, the contribution $S_2$ is formally expanded in integer powers of $1/N$.  The first term in the square brackets is the bulk action, while the second term is an endpoint correction.  The final term in the square brackets, along with the $N\log N$ term arises from the bulk, and can be traced to the log divergence when $\lambda_i$ approaches $\lambda_j$.  Note, however, that additional powers of $1/N^\alpha$ will be obtained when expanding the bulk action in the large-$N$ limit.

The leading order effective action, (\ref{eq:S0}), is obtained by noting that the function $f(x,\tilde x)$ in (\ref{eq:fxxt}) becomes highly peaked at $x\approx\tilde x$ in the large-$N$ limit.  Based on the form of this function, we make the substitution
\begin{equation}
    w=N^\alpha(1+iy'(x))(\tilde x-x).
\end{equation}
In addition, since the first term in (\ref{eq:S2N}) is integrated symmetrically in $x$ and $\tilde x$, we may consider the symmetrical combination $f(x,\tilde x)+f(\tilde x,x)$.  The expansion then takes the form
\begin{align}
    \rho(\tilde x)f_s(x,\tilde x)&=\rho f_s(w)+N^{-\alpha}\left[\fft{\rho'}{1+iy'}wf_s(w)+\fft{i}2\fft{\rho y''}{(1+iy')^2}w^2f_s'(w)\right]\nn\\
    &\quad+N^{-2\alpha}\left[\fft12\fft{\rho''}{(1+iy')^2}w^2f_s(w)+\fft{i}6\fft{3\rho'y''+\rho y'''}{(1+iy')^3}w^3f_s'(w)-\fft18\fft{\rho y''^2}{(1+iy')^4}w^4f_s''(w)\right].\nn\\
    &\quad+\mathcal O(N^{-3\alpha})
\end{align}
Here $f_s(z)=\fft12(f(z)+f(-z))$ where $f(z)$ is given in (\ref{eq:f(z)}) is explicitly symmetric in $z$, and we have suppressed the explicit $x$ dependence of the functions $\rho$ and $y$ for notational convenience.

Note that the change of variables from $\tilde x$ to $w$ leads to an integral of the form
\begin{equation}
    \int_{-x_*}^{x_*}d\tilde x=\fft{N^{-\alpha}}{1+iy'}\int_{-N^\alpha(1+iy')(x_*+x)}^{N^\alpha(1+iy')(x_*-x)}dw.
    \label{eq:intlim}
\end{equation}
As long as $x$ is not near the endpoints, $\pm x_*$, this integral can be extended to $\pm\infty$ since $f_s(w)$ vanishes exponentially for large arguments (assuming we do not cross any Stokes lines when deforming away from the real axis).  In this case, the integral over $w$ of the $N^{-\alpha}$ term vanishes because the integrand is odd.  For the other terms, we may use the definite integrals
\begin{equation}
    \int_{-\infty}^\infty f_s(w)dw=-\fft{16\pi^2}{27},\qquad\int_{-\infty}^\infty w^2f_s(w)dw=-\fft{32\pi^4}{243},
\end{equation}
along with integration by parts (with vanishing endpoints) to obtain an effective action
\begin{align}
    \mathcal F_0&=-\fft{ik}{4\pi}N^{1+2\alpha}\left[\int_{-x_*}^{x_*}dx\rho(x+iy)^2+\mathcal O(N^{-1})\right]\nn\\
    &\quad+\fft{16\pi^2}{27}N^{2-\alpha}\int dx\fft\rho{1+iy'}\Bigl[\rho+
    \fft{2\pi^2}9N^{-2\alpha}\left(\fft12\fft{\rho''}{(1+iy')^2}-\fft{i}2\fft{3\rho'y''+\rho y'''}{(1+iy')^3}-\fft32\fft{\rho y''^2}{(1+iy')^4}\right)\nn\\
    &\kern12em+\mathcal O(N^{-4\alpha})\Bigr]\nn\\
    &\quad-N\int_{-x_*}^{x_*}dx\rho(x)\log\left(\fft{2\pi\rho}{1+iy'}\right)-(1-\alpha)N\log N+(\mbox{endpoints})+\mathcal O(1),
    \label{eq:effa}
\end{align}
where we have included the $S_1$ term, (\ref{eq:S1}), obtained above.  Taking $\alpha=1/3$, the leading order contribution is at $\mathcal O(N^{5/3})$, and matches the expression (\ref{eq:S0}) obtained previously in \cite{Guarino:2015jca}.  More generally, we note that the large-$N$ expansion include competing powers of $N^{-\alpha}$ from the eigenvalues, (\ref{eq:lambdai}), and $N^{-1}$ from Euler-Maclaurin summation.

It should be noted that we have not included any endpoint corrections in the expression for the effective action, (\ref{eq:effa}).  At the order we are considering, these include both the second term in the square brackets of (\ref{eq:S2N}) and endpoint corrections when one of the limits of integration in (\ref{eq:intlim}) cannot be extended to infinity.  Since $f_s(z)$ is exponentially suppressed away from zero, the endpoint corrections are only important in a region of width $\mathcal O(N^{-\alpha})$ near the endpoints. This will have no effect on the leading order calculation of the free energy, but becomes important at subleading order.

\subsection{The eigenvalue distribution at subleading order}

Away from the endpoints, we can find the next order corrections to the eigenvalue density $\rho(x)$ and imaginary components $y(x)$ by varying the effective action (\ref{eq:effa}) with the inclusion of a Lagrange multiplier in order to enforce the constraint that $\rho(x)$ is properly normalized.  Taking $\alpha=1/3$, the leading order contribution to the action is of $\mathcal O(N^{5/3})$, and the first subleading correction is of $\mathcal O(N)$ and arises from a combination of the second and final lines of (\ref{eq:effa}).

As observed numerically, the first subleading corrections to $\rho(x)$ and $y(x)$ scale as $\mathcal O(N^{-2/3})$, which is consistent with the structure of (\ref{eq:effa}).  As a result, we can take a perturbative expansion
\begin{align}
\rho(x)&=\rho_0(x)+N^{-2/3}\rho_1(x)+\mathcal O(N^{-1}),\nn\\
y(x)&=y_0(x)+N^{-2/3}y_1(x)+\mathcal O(N^{-1}),
\end{align}
where $\rho_0(x)$ and $y_0(x)$ correspond to the leading order solution given in (\ref{eq:leading}).  Varying (\ref{eq:effa}) with respect to $\rho(x)$ and substituting in the leading order solution then gives
\begin{equation}
    \mu_1=\fft{k}{2\pi}(x+iy_0)y_1+\fft{32\pi^2}{27}\left[\fft{\rho_1}{1+iy_0'}-i\fft{\rho_0y_1'}{(1+iy_0')^2}+\fft{\pi^2}9\fft{\rho_0''}{(1+iy_0')^3}\right]-\log\left(\fft{2\pi\rho_0}{1+iy_0'}\right)-1,
    \label{eq:rho1eq}
\end{equation}
where the subleading Lagrange multiplier $\mu_1$ may be complex.  Note that this expression has already been simplified for $y_0(x)=x/\sqrt3$ being a linear function of $x$.

The equation (\ref{eq:rho1eq}) is in general a complex equation.  However, we demand the functions $\rho_1(x)$ and $y_1(x)$ to be real.  This is now sufficient for us to obtain the solution
\begin{align}
\rho_1(x)&=\frac{9}{16 \pi^2}\log\left(1-(x/x_*)^2\right)+C_1,\nonumber\\
y_1(x)&=-\fft{3\sqrt3}{8\pi^2}\frac{x\log\left(1-(x/x_*)^2\right)+2x_*\tanh^{-1}(x/x_*)+C_2x}{\rho_0(x)},
\label{eq:rho1}
\end{align}
where $C_1$ and $C_2$ are constants related to the Lagrange multiplier that we have been unable to fix without a better understanding of the endpoint corrections.  We note that the subleading corrections $\rho_1(x)$ and $y_1(x)$ match the results of the numerical calculations quite well (apart from the endpoints), as shown in Figure~\ref{fig:rhosub}.  In addition, we have checked that they are consistent with the second equation this is obtained by varying the effective action (\ref{eq:effa}) with respect to $y(x)$.

\subsection{Cancellation of the \texorpdfstring{$N\log N$}{N log(N)} term} 
As we have seen, the effective action, (\ref{eq:effa}), contains a term of the form $N\log N$, which is not observed numerically in the free energy. This suggests that it ought to be cancelled by a similar contribution from the one-loop determinant, (\ref{eq:F1det}).  We now demonstrate analytically that this is indeed what happens.  To do so, we start with the components of the Hessian matrix $B_{ij}=\partial^2\mathcal F_0/\partial\lambda_i\partial\lambda_j$
\begin{align}
    B_{ii}&=-\fft{ik}{2\pi}+\sum_{k\ne i}\left(\fft12\csch^2\fft{\lambda_{ik}}2+h'(\lambda_{ik})\right),\nn\\
    B_{ij}&=-\fft12\csch^2\fft{\lambda_{ij}}2-h'(\lambda_{ij})\qquad(i\ne j),
    \label{eq:BiiBij}
\end{align}
where
\begin{equation}
    h(z)=\fft{2\sinh z-\fft{3\sqrt3}{2\pi}z}{1+2\cosh z}.
\end{equation}
is a smooth function that is exponentially suppressed for large $z$.  The dominant contribution to the Hessian matrix comes from the $\csch^2(\lambda_{ij}/2)$ factors which are large on and near the diagonal.

In order to evaluate the determinant, we can break up the $B$ matrix into its diagonal and off-diagonal components $B=B_d+B_{od}=B_d(1+B_d^{-1}B_{od})$ so that
\begin{align}
    \log\det B&=\Tr\log B_d+\Tr\log(1+B_d^{-1}B_{od})\nn\\
    &=\sum_i\log B_{ii}+\sum_{n=1}^\infty\fft{(-1)^{n+1}}n\Tr(B_d^{-1}B_{od})^n,
\end{align}
where we have formally expanded the log.  Although $B_{od}$ is not necessarily small, the matrix $B_d^{-1}B_{od}$ obtained by scaling by the diagonal entries remains bounded.  Thus we expect that the determinant is dominated by the diagonal elements, and hence will focus only on the diagonal contribution.

In order to evaluate the diagonal elements $B_{ii}$, we convert the sum in (\ref{eq:BiiBij}) into an integral.  However, as in the evaluation of $S_2$ in (\ref{eq:S2reg}), we have to treat the $\lambda_i\to\lambda_j$ divergence with care.  In fact, we can apply the same regulation procedure as we did above by approximating $\lambda_{ik}$ by $(i-k)\alpha(i)$ and then writing
\begin{equation}
   B_{ii}=-\fft{ik}{2\pi}+\sum_{k\ne i}\fft2{(i-k)^2\alpha(i)^2}+\sum_k\left(\fft12\csch^2\fft{\lambda_{ik}}2+h'(\lambda_{ik})-\fft2{(i-k)^2\alpha(i)^2}\right)+\fft16.
\end{equation}
The factor of $1/6$ is introduced to cancel the contribution from $k=i$ in the unrestricted sum on the right-hand side.  Ignoring boundary effects, which lead to higher order corrections, we can extend the limits of the first sum to infinity and convert the second sum to an integral, with the result
\begin{equation}
     B_{ii}=-\fft{ik}{2\pi}+\fft{2\pi^2}{3\alpha(i)^2}+\int dj\left(\fft12\csch^2\fft{\lambda_{i+j}-\lambda_i}2+h'(\lambda_{i+j}-\lambda_i)-\fft2{j^2\alpha(i)^2}\right)+\mathcal O(1)
\end{equation}
Given an eigenvalue distribution specified by $\rho(x)$ and $y(x)$, we can convert the integral over the index $j$ into an integral over $x$.  However, to obtain the dominant $N\log N$ behavior, it is sufficient to make the approximation $\lambda_{i+j}-\lambda_i\approx j\alpha(i)$.  The integral can then be performed, with the result
\begin{equation}
    B_{ii}\approx\fft{2\pi^2}{3\alpha(i)^2}
    =N^{2-2\alpha}\fft{2\pi^2\rho(x)^2}{3(1+iy'(x))^2}+\cdots,
\end{equation}
where we substituted in $\alpha(i)$ from (\ref{eq:alphai}) and dropped the $-ik/2\pi$ term as it is subdominant in the large-$N$ limit.

The determinant contribution to the free energy is then
\begin{align}
    \mathcal F_1\approx\fft12\sum_i\log B_{ii}&\approx\fft{N}2\int_{-x_*}^{x_*} dx\rho\log\left(N^{2-2\alpha}\fft{2\pi^2\rho^2}{3(1+iy')^2}\right)\nn\\
    &=N\int_{-x_*}^{x_*} dx\rho\log\left(\fft{2\pi\rho}{1+iy'}\right)+(1-\alpha)N\log N-\fft{N}2\log 6+\cdots.
\end{align}
Comparison with (\ref{eq:effa}) demonstrates that not only the $N\log N$ term but also the integral term, which is linear in $N$, cancels similar contributions in the effective action.  (The cancellation at $\mathcal O(N)$ is not complete, however, as there is a $-(1/2)\log6$ term left over.)  Actually, all of these terms arise from the log divergence in $\mathcal F_0$ when $\lambda_i$ approaches $\lambda_j$, so it is perhaps not a surprise to see such a cancellation.  Of course, we have not yet examined the off-diagonal contribution to the determinant, which would be expected to contribute at higher orders (including at $\log N$ order), but would not spoil the leading $N\log N$ cancellation.

\section{Discussion}
\label{sec:discussion}

Our main result is numerical evidence for log contributions to the free energy of the form
\begin{equation}
    \Re F(N,k)=f_0 N^{5/3}k^{1/3}+\cdots+\fft29\log N-\fft1{18}\log k+\cdots.
\end{equation}
Ideally, we would like to obtain an analytic understanding of the $2/9$ and $1/18$ coefficients.  However, this has proven to be a challenge, as the expansion to subleading order requires particular care near the endpoints.  For example, as we have seen in (\ref{eq:rho1}), the eigenvalue density away from the endpoints receives a correction of $\mathcal O(N^{-2/3})$.  In contrast, the endpoint corrections start at $\mathcal O(1)$ at the endpoints, but fall off exponentially within a distance of $\mathcal O(N^{-1/3})$ from the endpoints.  Of course, coefficients in front of logs can sometimes be obtained without a full calculation, so there is still the possibility that a careful examination of the large-$N$ expansion including Euler-Maclaurin corrections can produce the log terms in the free energy.

Beyond the log terms, we have been able to match the structure of the `t~Hooft expansion up to genus-one.  Since we only compute the saddle point contribution and one-loop determinant, this is the limit of what we are able to probe numerically.  In principle, a full numerical analysis would go beyond a numerical saddle point evaluation.  (This was, for example, carried out using Monte Carlo integration in \cite{Hanada:2012si} for ABJM theory.)
However, as we were mainly in interested in exploration of the log terms, the numerical saddle point expansion is sufficient and allows us to work with $N$ up to 600 without too much difficulty.

Just as the free energies of ABJM-like theories with $N^{3/2}$ scaling have a universal contribution of the form $(1/4)\log N$ (where $k$ is kept fixed), we may expect theories with $N^{5/3}$ scaling to have a universal log contribution as well.  This leads us to conjecture that the $(2/9)\log N$ term that we obtained numerically is universal for a large class of Chern-Simons-matter theories dual to massive IIA theory.  This $2/9$ coefficient corresponds to the large-$N$ limit where the Chern-Simons level $k$ or levels $k_a$ are held fixed.

In the case of ABJM-like theories, the universal $(1/4)\log N$ behavior is easily obtained on the field theory side by writing the partition function as an Airy function \cite{Fuji:2011km,Marino:2011eh} and then taking the large-$N$ limit.  For theories with $N^{5/3}$ scaling, however, the general structure of the full partition function is not yet known.  Thus we do not have a similar justification for universality of the $\log N$ term.  Nevertheless, a basis for universality can be seen on the supergravity side of the duality.  The $(1/4)\log N$ behavior of ABJM-like theories can be obtained by a universal one-loop calculation in 11-dimensional supergravity \cite{Bhattacharyya:2012ye}, and we suggest a similar argument can be made for universality of the one-loop log term in massive IIA theory.  This is not entirely straightforward, however, as the log term only arises from zero modes in 11-dimensional supergravity, but could arise more generally in the non-zero-mode part of a 10-dimensional heat kernel calculation.  Thus it would certainly be worthwhile to perform a one-loop massive IIA calculation, both as a test of precision holography and as an indicator of universality of log corrections to the partition function.

\acknowledgments
We wish to thank L.~Pando~Zayas for illuminating discussions on the subleading structure of supersymmetric partition functions, especially on the nature of $\log N$ versus $\log\lambda$ corrections and on the origin of terms in the exact partition function that are not captured by the genus expansion.  This work was supported in part by the U.S.~Department of Energy under grant DE-SC0007859.

\bibliographystyle{JHEP}
\bibliography{massive-refs}

\providecommand{\href}[2]{#2}\begingroup\raggedright\begin{thebibliography}{10}

\bibitem{Aharony:2008ug}
O.~Aharony, O.~Bergman, D.~L. Jafferis and J.~Maldacena, \emph{{$\mathcal N=6$
  superconformal Chern-Simons-matter theories, M2-branes and their gravity
  duals}}, \href{https://doi.org/10.1088/1126-6708/2008/10/091}{\emph{JHEP}
  {\bfseries 10} (2008) 091} [\href{https://arxiv.org/abs/0806.1218}{{\ttfamily
  0806.1218}}].

\bibitem{Gaiotto:2009mv}
D.~Gaiotto and A.~Tomasiello, \emph{{The gauge dual of Romans mass}},
  \href{https://doi.org/10.1007/JHEP01(2010)015}{\emph{JHEP} {\bfseries 01}
  (2010) 015} [\href{https://arxiv.org/abs/0901.0969}{{\ttfamily 0901.0969}}].

\bibitem{Gaiotto:2009yz}
D.~Gaiotto and A.~Tomasiello, \emph{{Perturbing gauge/gravity duals by a Romans
  mass}}, \href{https://doi.org/10.1088/1751-8113/42/46/465205}{\emph{J. Phys.}
  {\bfseries A42} (2009) 465205}
  [\href{https://arxiv.org/abs/0904.3959}{{\ttfamily 0904.3959}}].

\bibitem{Marino:2009jd}
M.~Marino and P.~Putrov, \emph{{Exact Results in ABJM Theory from Topological
  Strings}}, \href{https://doi.org/10.1007/JHEP06(2010)011}{\emph{JHEP}
  {\bfseries 06} (2010) 011} [\href{https://arxiv.org/abs/0912.3074}{{\ttfamily
  0912.3074}}].

\bibitem{Drukker:2010nc}
N.~Drukker, M.~Marino and P.~Putrov, \emph{{From weak to strong coupling in
  ABJM theory}}, \href{https://doi.org/10.1007/s00220-011-1253-6}{\emph{Commun.
  Math. Phys.} {\bfseries 306} (2011) 511}
  [\href{https://arxiv.org/abs/1007.3837}{{\ttfamily 1007.3837}}].

\bibitem{Herzog:2010hf}
C.~P. Herzog, I.~R. Klebanov, S.~S. Pufu and T.~Tesileanu, \emph{{Multi-Matrix
  Models and Tri-Sasaki Einstein Spaces}},
  \href{https://doi.org/10.1103/PhysRevD.83.046001}{\emph{Phys. Rev.}
  {\bfseries D83} (2011) 046001}
  [\href{https://arxiv.org/abs/1011.5487}{{\ttfamily 1011.5487}}].

\bibitem{Drukker:2011zy}
N.~Drukker, M.~Marino and P.~Putrov, \emph{{Nonperturbative aspects of ABJM
  theory}}, \href{https://doi.org/10.1007/JHEP11(2011)141}{\emph{JHEP}
  {\bfseries 11} (2011) 141} [\href{https://arxiv.org/abs/1103.4844}{{\ttfamily
  1103.4844}}].

\bibitem{Fuji:2011km}
H.~Fuji, S.~Hirano and S.~Moriyama, \emph{{Summing Up All Genus Free Energy of
  ABJM Matrix Model}},
  \href{https://doi.org/10.1007/JHEP08(2011)001}{\emph{JHEP} {\bfseries 08}
  (2011) 001} [\href{https://arxiv.org/abs/1106.4631}{{\ttfamily 1106.4631}}].

\bibitem{Marino:2011eh}
M.~Marino and P.~Putrov, \emph{{ABJM theory as a Fermi gas}},
  \href{https://doi.org/10.1088/1742-5468/2012/03/P03001}{\emph{J. Stat. Mech.}
  {\bfseries 1203} (2012) P03001}
  [\href{https://arxiv.org/abs/1110.4066}{{\ttfamily 1110.4066}}].

\bibitem{Bhattacharyya:2012ye}
S.~Bhattacharyya, A.~Grassi, M.~Marino and A.~Sen, \emph{{A One-Loop Test of
  Quantum Supergravity}},
  \href{https://doi.org/10.1088/0264-9381/31/1/015012}{\emph{Class. Quant.
  Grav.} {\bfseries 31} (2014) 015012}
  [\href{https://arxiv.org/abs/1210.6057}{{\ttfamily 1210.6057}}].

\bibitem{Benini:2015noa}
F.~Benini and A.~Zaffaroni, \emph{{A topologically twisted index for
  three-dimensional supersymmetric theories}},
  \href{https://doi.org/10.1007/JHEP07(2015)127}{\emph{JHEP} {\bfseries 07}
  (2015) 127} [\href{https://arxiv.org/abs/1504.03698}{{\ttfamily
  1504.03698}}].

\bibitem{Benini:2015eyy}
F.~Benini, K.~Hristov and A.~Zaffaroni, \emph{{Black hole microstates in
  AdS$_{4}$ from supersymmetric localization}},
  \href{https://doi.org/10.1007/JHEP05(2016)054}{\emph{JHEP} {\bfseries 05}
  (2016) 054} [\href{https://arxiv.org/abs/1511.04085}{{\ttfamily
  1511.04085}}].

\bibitem{Liu:2017vll}
J.~T. Liu, L.~A. Pando~Zayas, V.~Rathee and W.~Zhao, \emph{{Toward Microstate
  Counting Beyond Large $N$ in Localization and the Dual One-loop Quantum
  Supergravity}}, \href{https://doi.org/10.1007/JHEP01(2018)026}{\emph{JHEP}
  {\bfseries 01} (2018) 026}
  [\href{https://arxiv.org/abs/1707.04197}{{\ttfamily 1707.04197}}].

\bibitem{PandoZayas:2019hdb}
L.~A. Pando~Zayas and Y.~Xin, \emph{{The Topologically Twisted Index in the 't
  Hooft Limit and the Dual AdS$_4$ Black Hole Entropy}},
  \href{https://arxiv.org/abs/1908.01194}{{\ttfamily 1908.01194}}.

\bibitem{Jeon:2017aif}
I.~Jeon and S.~Lal, \emph{{Logarithmic Corrections to Entropy of Magnetically
  Charged AdS$_4$ Black Holes}},
  \href{https://doi.org/10.1016/j.physletb.2017.09.026}{\emph{Phys. Lett.}
  {\bfseries B774} (2017) 41}
  [\href{https://arxiv.org/abs/1707.04208}{{\ttfamily 1707.04208}}].

\bibitem{Liu:2017vbl}
J.~T. Liu, L.~A. Pando~Zayas, V.~Rathee and W.~Zhao, \emph{{One-Loop Test of
  Quantum Black Holes in anti-de Sitter Space}},
  \href{https://doi.org/10.1103/PhysRevLett.120.221602}{\emph{Phys. Rev. Lett.}
  {\bfseries 120} (2018) 221602}
  [\href{https://arxiv.org/abs/1711.01076}{{\ttfamily 1711.01076}}].

\bibitem{Guarino:2015jca}
A.~Guarino, D.~L. Jafferis and O.~Varela, \emph{{String Theory Origin of Dyonic
  $\mathcal N=8$ Supergravity and Its Chern-Simons Duals}},
  \href{https://doi.org/10.1103/PhysRevLett.115.091601}{\emph{Phys. Rev. Lett.}
  {\bfseries 115} (2015) 091601}
  [\href{https://arxiv.org/abs/1504.08009}{{\ttfamily 1504.08009}}].

\bibitem{Hosseini:2017fjo}
S.~M. Hosseini, K.~Hristov and A.~Passias, \emph{{Holographic microstate
  counting for AdS$_{4}$ black holes in massive IIA supergravity}},
  \href{https://doi.org/10.1007/JHEP10(2017)190}{\emph{JHEP} {\bfseries 10}
  (2017) 190} [\href{https://arxiv.org/abs/1707.06884}{{\ttfamily
  1707.06884}}].

\bibitem{Benini:2017oxt}
F.~Benini, H.~Khachatryan and P.~Milan, \emph{{Black hole entropy in massive
  Type IIA}}, \href{https://doi.org/10.1088/1361-6382/aa9f5b}{\emph{Class.
  Quant. Grav.} {\bfseries 35} (2018) 035004}
  [\href{https://arxiv.org/abs/1707.06886}{{\ttfamily 1707.06886}}].

\bibitem{Azzurli:2017kxo}
F.~Azzurli, N.~Bobev, P.~M. Crichigno, V.~S. Min and A.~Zaffaroni, \emph{{A
  universal counting of black hole microstates in AdS$_{4}$}},
  \href{https://doi.org/10.1007/JHEP02(2018)054}{\emph{JHEP} {\bfseries 02}
  (2018) 054} [\href{https://arxiv.org/abs/1707.04257}{{\ttfamily
  1707.04257}}].

\bibitem{Liu:2018bac}
J.~T. Liu, L.~A. Pando~Zayas and S.~Zhou, \emph{{Subleading Microstate Counting
  in the Dual to Massive Type IIA}},
  \href{https://arxiv.org/abs/1808.10445}{{\ttfamily 1808.10445}}.

\bibitem{Kapustin:2009kz}
A.~Kapustin, B.~Willett and I.~Yaakov, \emph{{Exact Results for Wilson Loops in
  Superconformal Chern-Simons Theories with Matter}},
  \href{https://doi.org/10.1007/JHEP03(2010)089}{\emph{JHEP} {\bfseries 03}
  (2010) 089} [\href{https://arxiv.org/abs/0909.4559}{{\ttfamily 0909.4559}}].

\bibitem{Jafferis:2010un}
D.~L. Jafferis, \emph{{The Exact Superconformal R-Symmetry Extremizes Z}},
  \href{https://doi.org/10.1007/JHEP05(2012)159}{\emph{JHEP} {\bfseries 05}
  (2012) 159} [\href{https://arxiv.org/abs/1012.3210}{{\ttfamily 1012.3210}}].

\bibitem{Hama:2010av}
N.~Hama, K.~Hosomichi and S.~Lee, \emph{{Notes on SUSY Gauge Theories on
  Three-Sphere}}, \href{https://doi.org/10.1007/JHEP03(2011)127}{\emph{JHEP}
  {\bfseries 03} (2011) 127} [\href{https://arxiv.org/abs/1012.3512}{{\ttfamily
  1012.3512}}].

\bibitem{Jafferis:2011zi}
D.~L. Jafferis, I.~R. Klebanov, S.~S. Pufu and B.~R. Safdi, \emph{{Towards the
  F-Theorem: $\mathcal N=2$ Field Theories on the Three-Sphere}},
  \href{https://doi.org/10.1007/JHEP06(2011)102}{\emph{JHEP} {\bfseries 06}
  (2011) 102} [\href{https://arxiv.org/abs/1103.1181}{{\ttfamily 1103.1181}}].

\bibitem{Benini:2016hjo}
F.~Benini and A.~Zaffaroni, \emph{{Supersymmetric partition functions on
  Riemann surfaces}}, {\emph{Proc. Symp. Pure Math.} {\bfseries 96} (2017) 13}
  [\href{https://arxiv.org/abs/1605.06120}{{\ttfamily 1605.06120}}].

\bibitem{Benini:2018mlo}
F.~Benini and P.~Milan, \emph{{A Bethe Ansatz type formula for the
  superconformal index}},  \href{https://arxiv.org/abs/1811.04107}{{\ttfamily
  1811.04107}}.

\bibitem{Ooguri:2002gx}
H.~Ooguri and C.~Vafa, \emph{{World sheet derivation of a large $N$ duality}},
  \href{https://doi.org/10.1016/S0550-3213(02)00620-X}{\emph{Nucl. Phys.}
  {\bfseries B641} (2002) 3}
  [\href{https://arxiv.org/abs/hep-th/0205297}{{\ttfamily hep-th/0205297}}].

\bibitem{Hanada:2012si}
M.~Hanada, M.~Honda, Y.~Honma, J.~Nishimura, S.~Shiba and Y.~Yoshida,
  \emph{{Numerical studies of the ABJM theory for arbitrary $N$ at arbitrary
  coupling constant}},
  \href{https://doi.org/10.1007/JHEP05(2012)121}{\emph{JHEP} {\bfseries 05}
  (2012) 121} [\href{https://arxiv.org/abs/1202.5300}{{\ttfamily 1202.5300}}].

\bibitem{Hatsuda:2012dt}
Y.~Hatsuda, S.~Moriyama and K.~Okuyama, \emph{{Instanton Effects in ABJM Theory
  from Fermi Gas Approach}},
  \href{https://doi.org/10.1007/JHEP01(2013)158}{\emph{JHEP} {\bfseries 01}
  (2013) 158} [\href{https://arxiv.org/abs/1211.1251}{{\ttfamily 1211.1251}}].

\bibitem{Calvo:2012du}
F.~Calvo and M.~Marino, \emph{{Membrane instantons from a semiclassical TBA}},
  \href{https://doi.org/10.1007/JHEP05(2013)006}{\emph{JHEP} {\bfseries 05}
  (2013) 006} [\href{https://arxiv.org/abs/1212.5118}{{\ttfamily 1212.5118}}].

\bibitem{Hatsuda:2013oxa}
Y.~Hatsuda, M.~Marino, S.~Moriyama and K.~Okuyama, \emph{{Non-perturbative
  effects and the refined topological string}},
  \href{https://doi.org/10.1007/JHEP09(2014)168}{\emph{JHEP} {\bfseries 09}
  (2014) 168} [\href{https://arxiv.org/abs/1306.1734}{{\ttfamily 1306.1734}}].

\end{thebibliography}\endgroup

\end{document}